\documentstyle[12pt]{article}
\def\a{\alpha}
\def\b{\beta}
\def\c{\chi}
\def\d{\delta}
\def\e{\epsilon}

\def\g{\gamma}

\def\p{\psi}
\def\k{\kappa}
\def\l{\lambda}
\def\m{\mu}

\def\o{\omega}
\def\t{\theta}
\def\r{\rho}
\def\s{\sigma}

\def\z{\zeta}

\def\vf{\varphi}

\def\bz{\bar{z}}

\def\bp{\bar{\p}}

\def\be{\begin{equation}}
\def\ee{\end{equation}}
\def\arr{\begin{array}{rll}}
\def\ea{\end{array}}
\def\bea{\begin{eqnarray}}
\def\eea{\end{eqnarray}}

\begin{document}

\begin{titlepage}
\noindent
\begin{flushright}
LNF--01/010(P) \\
ITP--UH--02/01 
\end{flushright}

\vskip 1.5cm

\begin{center}

{\Large\bf A Heterotic N=2 String}\\

\vskip 0.3cm

{\Large\bf with Space--Time Supersymmetry}\\

\bigskip

\vskip 1.5cm

{\large Stefano Bellucci}~\footnote{
E-mail: Stefano.Bellucci@lnf.infn.it}\ ,\ \
{\large Anton Galajinsky}~\footnote{
On leave from Department of Mathematical Physics,
Tomsk Polytechnical University, Tomsk, Russia\\
\phantom{XX}   
E-mail: Anton.Galajinsky@lnf.infn.it}

\vskip 0.4cm

{\it INFN--Laboratori Nazionali di Frascati, C.P. 13, 
00044 Frascati, Italia}\\

\vskip 0.7cm

and

\vskip 0.7cm
 
{\large Olaf Lechtenfeld}~\footnote{
E-mail: lechtenf@itp.uni-hannover.de}

\vskip 0.4cm

{\it Institut f\"ur Theoretische Physik, Universit\"at Hannover,\\
Appelstra\ss e 2, 30167 Hannover, Germany}

\end{center}

\vskip 1.5cm

\begin{abstract}
\noindent
We reconsider the issue of embedding space--time fermions into the
four-dimen\-sion\-al $N{=}2$ world--sheet supersymmetric string. 
A new heterotic theory is constructed, taking the right--movers from 
the $N{=}4$ topological extension of the conventional $N{=}2$ string 
but a $c{=}0$ conformal field theory supporting target--space supersymmetry 
for the left--moving sector. The global bosonic symmetry of the full formalism 
proves to be $U(1,1)$, just as in the usual $N{=}2$ string. 
Quantization reveals a spectrum of only two physical states,
one boson and one fermion, which fall in a multiplet of $(1,0)$ supersymmetry.

\end{abstract}

\vspace{0.5cm}


\end{titlepage}

\noindent
{\bf 1. Introduction}\\[-4pt]

\noindent
The $N{=}2$ string has attracted considerable interest over the past decade
(see \cite{marcus} for an older review and \cite{bischoff} for BRST 
quantization). Being relatively simple to analyze, 
the theory displays a few peculiar properties 
which distinguish it among others. To mention only the most significant 
points, there appears only a finite number of physical states in the 
excitation spectrum. However, one
encounters a continuous family of sectors interpolating between $R$ and 
$NS$ and being connected by spectral flow~\cite{schwimmer}. The latter 
is essentially a consequence of the complex geometry intrinsic to the 
RNS formulation of the $N{=}2$ string (see e.g.~\cite{kiritsis}). 
A remnant of the gauged $U(1)$ R~symmetry, spectral flow enables one to 
restrict oneself to a preferred sector, say the NS one.
All tree--level amplitudes with more 
than three external legs prove to vanish~\cite{ov1, ov} (for details of
loop calculations  see a recent work~\cite{cln} and references therein),
in complete agreement with the fact that the massless excitations of the 
$N{=}2$ string parametrize self--dual gauge or gravitational theory
in two spatial and two temporal dimensions~\cite{ov}. 
Also worth mentioning is a ``universal string'' interpretation 
deriving from the fact that $N{=}0$ and $N{=}1$ strings can be 
viewed as special classes of vacua of the $N{=}2$ string~\cite{berkvafa1}.

On the downside, the Brink-Schwarz action~\cite{brink} lacks
manifest $SO(2,2)$ Lorentz covariance, and the world--sheet theory
of $N{=}2$ supergravity coupled to matter supermultiplets
is unable to produce space--time fermions in the excitation spectrum.
An early attempt~\cite{klp} to gain fermions by adding
twisted sectors~\cite{mm1} to the $N{=}2$ string failed because the
necessary GSO projection prevented any interaction between
bosonic and fermionic states 
(see, however, \cite{berksiegel} for a different strategy).

Alternatively, a manifestly space--time
supersymmetric and $SO(2,2)$ covariant Green--Schwarz--type formulation 
of the $N{=}2$ string was proposed in~\cite{siegel}. Yet, as was 
later recognized~\cite{pope1,pope2}, the set of currents invented 
in~\cite{siegel} does not form 
a closed algebra. This drawback was overcome in~\cite{pope1,pope2} 
by taking a smaller but closed subset of the constraints. 
However, the massless states of the resulting $N{=}2$ string, 
although consisting of a scalar and a spinor, do not interact according to 
self--dual super Yang--Mills or self--dual supergravity.

A more successful approach has been advocated by de Boer and 
Skenderis~\cite{deboer} who combined a right--moving heterotic $N{=}(1,2)$ 
string with a left--moving Green--Schwarz--Berkovits type sigma 
model~\cite{gsb}. Since, by the very construction, the
$N{=}(1,2)$ string lives in a two- or three-dimensional target, the analysis
of~\cite{deboer} is likely to yield only a dimensional reduction of 
self--dual supergravity to $2{+}1$ dimensions. A common feature of both
approaches is the introduction of momenta canonically conjugate 
to the space--time spinors and a manifest Lorentz covariance.
 
Quite recently, the zero--mode structure was investigated for
a potential string theory which would be capable of describing 
supersymmetric self--dual Yang--Mills in $2{+}2$ dimensions~\cite{gl}. 
In agreement with~\cite{deboer} the analysis suggests that 
the ultimate formulation seems to be doubly supersymmetric, i.e. 
possessing both local world--sheet supersymmetry and kappa symmetry.

In the present paper we reconsider the issue of embedding space--time 
fermions into the $N{=}2$ string. We take advantage of the 
existing literature~\cite{pope1,deboer} and construct a new heterotic
string in two spatial and two temporal dimensions. 
In our scenario self--duality is implemented by the right--movers while 
manifest target--space supersymmetry is captured by left--movers.
The drawbacks of the previous attempts we discussed above suggest, however, 
that the presence of space--time fermions in the spectrum might be
incompatible with manifest $SO(2,2)$ Lorentz invariance.
Here, we choose to relax the latter property and build a new heterotic
string which supports $U(1,1)$, just as the conventional $N{=}2$ string does. 

Since there are two time--like directions in the target space
one can introduce two light--cone structures. When specifying left--moving
degrees of freedom we use the hitherto unexploited possibility to 
supersymmetrize only one light--cone direction just in the way it works in 
$(1,0)$ supergravity (see e.g. Ref.~\cite{elo}). 
Remarkably, this can be done without spoiling the 
global $U(1,1)$ group which is installed in the formalism by the left--movers. 
The system of currents we use here looks very similar to that
examined in~\cite{pope1} but differs in the global symmetry structure.
As to the right--movers, a first guess would be to take those of the 
conventional $N{=}2$ string. Surprisingly, this proves to be incompatible with 
the $U(1,1)$ kept by the left--movers. In order to reconcile the two points  
we turn to the $N{=}4$ topological reformulation of the $N{=}2$ string 
proposed by Berkovits and Vafa~\cite{berkvafa} and further studied 
in~\cite{bvw}, which thus specifies the right--moving sector of the model.
Quantization of the theory reveals only two physical states in the spectrum, 
one boson and one fermion, which prove to fall in a multiplet
of the $(1,0)$ space--time supersymmetry. 

The organization of the paper is as follows. In Sect.~2 we briefly review the 
salient features concerning the conventional $N{=}2$ string and its $N{=}4$ 
topological reformulation. We also specify the right--movers and fix our 
notation. Sect.~3 contains a description of the left--moving sector and a 
realization of the space--time supersymmetry. The global symmetry structure of 
the complete model is discussed in detail in Sect.~4. In particular, 
a supersymmetric extension of the $U(1,1)$ group is given. Sects.~5 and~6 
perform the quantization of the right- and left--movers, respectively. 
We summarize our results and discuss possible further developments in Sect.~7. 

\vspace{0.6cm}

\noindent
{\bf 2. The right--moving sector}\\[-4pt]

\noindent
As has been discussed in the Introduction, the right--movers 
in the new model are designed to keep the self--duality for the massless
states. Guided by the formalism for the conventional closed $N{=}2$ string, 
it seems natural to choose them to be just the right--movers of the latter. 
We proceed directly to the superconformal gauge. 
Then the space of histories is spanned by the canonical 
pair $(z^a,p_{z a})$, $a=0,1$, its complex conjugate $({\bar z}^a, 
p_{\bar z  a})$, and a couple of complex conjugate fermions 
$\psi^ a, {\bar\psi}^a$. The canonical brackets imposed on the fields are
\bea\label{brac}
&& \{ z^a (\s), p_{z}^b (\s') \}=\eta^{ab} \delta(\s- \s'), \quad
\{ {\bar z}^a (\s), p_{\bar z}^b (\s') \}=\eta^{ab} \delta(\s- \s'), 
\nonumber\\[2pt]
&&\{ \p^a (\s), {\bp}^b (\s') \}=i\eta^{ab } \delta(\s- \s'),
\eea
with the Minkowski metric $\eta^{ab}={\it diag} (-,+).$

The dynamics in the sector is governed by the Hamiltonian
\be
H=\int^{2\pi}_{0}\!\! d \s \{ 2\pi(p_z^a p_{\bz a} +{\textstyle{\frac {1}
{{(2\pi)}^2}}} \partial_1 z^a \partial_1 \bz_a)-{\textstyle{\frac i2}}
{\bp}^a \partial_1 \p_a
-{\textstyle{\frac i2}} {\p}^a \partial_1 {\bp}_a \},
\ee
where $\partial_1$ denotes the derivative with respect to $\s$.

Given the space, one can build a 
set of currents (a contraction of indices is implied),
\bea\label{n2}
&&T=(p_z +{\textstyle{\frac {1}{2\pi}}} \partial_1 \bz)
(p_{\bz} +{\textstyle{\frac {1}{2\pi}}} \partial_1 z)-
{\textstyle{\frac {i}{2\pi}}}(\p \partial_1 \bp +
\bp \partial_1 \p)=0,\nonumber\\[2pt]
&& G=(p_z +{\textstyle{\frac {1}{2\pi}}} \partial_1 \bz) \p =0,
\qquad \bar G=(p_{\bz} +{\textstyle{\frac {1}{2\pi}}} 
\partial_1 z) \bp =0,\nonumber\\[2pt]
&& J =\bp \p =0,     
\eea
which is closed under the bracket~(\ref{brac}) and forms nothing but
the $N{=}2$ superconformal algebra.

Making use of the brackets~(\ref{brac}), the equations of motion can be 
easily derived from the Hamiltonian (as usual $\partial_{\pm}=
\partial_0  \pm\partial_1$),
\bea\label{eqm}
&&{\dot z}^a -2\pi {p_{\bz}}^a =0, \qquad 
{{\dot p}_{\bz}}^a -{\textstyle{\frac {1}{2\pi}}} \partial_1 
\partial_1 z^a=0,\nonumber\\[2pt]
&& {\dot{\bz}}^a -2\pi {p_z}^a =0, \qquad 
{{\dot p}_z}^a -{\textstyle{\frac {1}{2\pi}}} \partial_1 
\partial_1 {\bz}^a=0,\nonumber\\[2pt]
&& \partial_{-} \p^a =0, \qquad \partial_{-} {\bp}^a =0.
\eea
Being applied to the currents, these yield
\be\label{eqcur}
\partial_{-} T=0, \quad \partial_{-} G=0, \quad \partial_{-} \bar G =0, \quad
\partial_{-} J=0.
\ee

Notice further that the equations~(\ref{eqm}) are amended by the periodicity 
conditions
\be\label{bc}
(\delta z \partial_1 \bz +\delta \bz \partial_1 z) |^{2\pi}_{0} =0, \qquad
(\bp \delta \p + \p \delta \bp) |^{2\pi}_{0} =0.
\ee
The bosons describe string coordinates and are single--valued (periodic) in
the flat space. Due to their complex nature, the NSR fermions live in a 
twisted spinor bundle~\cite{mm}, i.e.
\bea\label{boundcond}
&& \p (\s+2\pi)=e^{2i\nu\pi} \p(\s), \qquad \bp(\s+2\pi)=e^{-2i\nu\pi} \bp(\s),
\eea
with $\nu$ being an arbitrary real number. For the currents this amounts to
\bea
G(\s+2\pi)=e^{2i\nu \pi} G(\s), \qquad 
\bar G (\s+2\pi)=e^{-2i\nu \pi} \bar G(\s), \nonumber\\[2pt] 
T(\s+2\pi)=T(\s), \qquad J(\s+2\pi)=J(\s).
\eea
The choice of the parameter~$\nu$ determines the moding for the fermionic
fields. 
This is in contrast with the conventional $N{=}1$ string for which only 
two options ($R$ or $NS$) are available.

Varying~$\nu$ changes the representation of the $N{=}2$
superconformal algebra, an effect known as ``spectral flow''.
However, all such representations are in fact equivalent~\cite{schwimmer}
Actually, it is straightforward to check that the continuous automorphism
(preserving Eq.~(\ref{eqcur})) 
\be
G'=e^{-i\a (\tau+\s)} G, \quad {\bar G'}=e^{i\a (\tau+\s)} \bar G, \quad
T'=T-{\textstyle{\frac {\a}{\pi}}}J, \quad J'=J, 
\ee
does the job for an arbitrary real number $\a$, since
\be
G'(\s+2\pi)=e^{2i(\nu-\a)\pi} G'(\s).
\ee
Because $\a$ is at our disposal, one can always stick with a preferred 
representation.\footnote{
Because $\a$ is a $U(1)$ modulus, 
it must be considered beyond tree level~\cite{cln}.}
For the rest of this paper we choose to work in the NS 
picture, thus just putting $\nu={\textstyle{\frac 12}}$ in 
Eq.~(\ref{boundcond}) above. 

It is worth noting further that, due to the complex structure
intrinsic to the $N{=}2$ string, the (spin cover of the) full target--space
Lorentz group $Spin(2,2)=SU(1,1)\times SU(1,1)'$ gets
broken to $U(1)\times SU(1,1)\simeq U(1,1)$ which is then the global 
symmetry group of the formalism under consideration. 

So far, our discussion made use of complex field variables.
In the next section, where we shall introduce the left--movers, it will 
turn out that the formulation is more transparent in a specific real 
field basis. We thus devote the remnant of this section to conform 
the present analysis to a real notation.
 
Given a fermionic field $\p^a$, with $a{=}0,1$, 
we first transform the vector index~$a$ to a ``light--cone'' basis,
\be
\p^{\pm}={\textstyle{\frac {1}{\sqrt 2}}}(\p^0 \pm \p^1), \quad
{\bp}^{\pm}={\textstyle{\frac {1}{\sqrt 2}}}({\bp}^0 \pm {\bp}^1),
\ee
and then take the real and imaginary parts,
\be
\p^{\pm}=\vf^{\pm} +i\c^{\pm}, \quad {\bar\p}^{\pm}=\vf^{\pm} -i\c^{\pm}, 
\ee
to be the new field variables. The only nonvanishing brackets are
\be\label{fermbrac}
\{ \vf^{+}(\s),\vf^{-}(\s') \}=-{\textstyle{\frac i2}} \delta(\s-\s'),
\quad \{ \c^{+}(\s),\c^{-}(\s') \}=-{\textstyle{\frac i2}} \delta(\s-\s').
\ee
Analogously, for the bosonic variables one performs a similar 
coordinate change,
\be
z^{\pm}={\textstyle{\frac {1}{\sqrt 2}}}(z^0 \pm z^1), \quad
{\bz}^{\pm}={\textstyle{\frac {1}{\sqrt 2}}}({\bz}^0 \pm {\bz}^1),
\ee
and defines the new fields
\be
z^{\pm}=X^{\pm} +iY^{\pm}, \quad p_z^{\pm}=p^{\pm} +i p_Y^{\pm},
\quad \bz^{\pm}=X^{\pm} -iY^{\pm}, \quad  
p_{\bz}^{\pm}=p^{\pm} -i p_Y^{\pm}.
\ee
The latter prove to obey
\bea\label{br1}
&&\{ X^{+} (\s),p^{-}(\s') \}=-{\textstyle{\frac 12}} \delta(\s-\s'),\quad
\{ X^{-} (\s),p^{+}(\s') \}=-{\textstyle{\frac 12}} \delta(\s-\s'),
\nonumber\\[2pt]
&&\{ Y^{+} (\s),p_Y^{-}(\s') \}={\textstyle{\frac 12}} \delta(\s-\s'),\quad
\{ Y^{-} (\s),p_Y^{+}(\s') \}={\textstyle{\frac 12}} \delta(\s-\s').
\eea

It is now trivial to conform the Hamiltonian 
to the new notation,
\bea\label{ham}
&&H=\int^{2\pi}_{0}\!\! d\s \{ -4\pi p^{+} p^{-}  -4\pi p_Y^{+} p_Y^{-}
-{\textstyle{\frac {1}{\pi}}} \partial_1 X^{+} \partial_1 X^{-}
-{\textstyle{\frac {1}{\pi}}} \partial_1 Y^{+} \partial_1 Y^{-}+
i\vf^{+} \partial_1 \vf^{-}+
\nonumber\\[2pt]
&& \quad \quad  i\vf^{-} \partial_1 \vf^{+}
+i\c^{+} \partial_1 \c^{-} +i\c^{-} \partial_1 \c^{+} \}.
\eea
Analogous manipulations with the set of currents, after some simple algebra, 
yield
\bea\label{car1}
&&\tilde T=-2(p^{-} +{\textstyle{\frac {1}{2\pi}}} \partial_1 X^{-})
(p^{+} +{\textstyle{\frac {1}{2\pi}}} \partial_1 X^{+})-2(p_Y^{-} -
{\textstyle{\frac {1}{2\pi}}} \partial_1 Y^{-})
(p_Y^{+} -{\textstyle{\frac {1}{2\pi}}} \partial_1 Y^{+})+\nonumber\\[2pt]
&&\quad \quad {\textstyle{\frac {i}{\pi}}}\vf^{+} \partial_1 \vf^{-}+
{\textstyle{\frac {i}{\pi}}}\vf^{-} \partial_1 \vf^{+}+
{\textstyle{\frac {i}{\pi}}}\c^{+} \partial_1 \c^{-} 
+{\textstyle{\frac {i}{\pi}}}\c^{-} \partial_1 \c^{+} =0,\nonumber\\[2pt]
&& \tilde G=(p^{+} +{\textstyle{\frac {1}{2\pi}}} \partial_1 X^{+}) \vf^{-}+
(p^{-} +{\textstyle{\frac {1}{2\pi}}} \partial_1 X^{-}) \vf^{+}-
(p_Y^{+} -{\textstyle{\frac {1}{2\pi}}} \partial_1 Y^{+})\c^{-}-
\nonumber\\[2pt]
&& \quad \quad (p_Y^{-} -{\textstyle{\frac {1}{2\pi}}} 
\partial_1 Y^{-})\c^{+}=0,
\nonumber\\[2pt]
&& \tilde H=(p^{+} +{\textstyle{\frac {1}{2\pi}}} \partial_1 X^{+}) \c^{-}+
(p^{-} +{\textstyle{\frac {1}{2\pi}}} \partial_1 X^{-}) \c^{+}+
(p_Y^{+} -{\textstyle{\frac {1}{2\pi}}} \partial_1 Y^{+})\vf^{-}+
\nonumber\\[2pt]
&& \quad \quad (p_Y^{-} -{\textstyle{\frac {1}{2\pi}}} 
\partial_1 Y^{-})\vf^{+}=0,\nonumber\\[2pt]
&& \tilde J =i\c^{+}\vf^{-}+i\c^{-}\vf^{+}=0.     
\eea
We do not see the vector indices any more, and the invariance under 
$U(1,1)$ is kept by contracting a ``${+}$`` with a ``${-}$`` 
(for more details see Sect.~4). The only nonvanishing brackets in practical
calculations are those involving plus and minus indices simultaneously.
In view of implementing global supersymmetry in the target, one
can now see an intriguing possibility: One may try to 
supersymmetrize in only {\it half\/} the spatial directions, 
making recourse to $(1,0)$ superspace (see e.g. Ref.~\cite{elo})! 
Thus, our main reason to prefer a real notation is the greater transparency
of the target--space supersymmetry in this basis.
 
No work has to be done with the equations of motion, these just acquiring 
the form
\bea
&&{\dot X}^{\pm} -2\pi p^{\pm} =0, \quad {\dot p}^{\pm} -{\textstyle
{\frac {1}{2\pi}}} \partial_1 \partial_1 X^{\pm} =0, \nonumber\\[2pt]
&&{\dot Y}^{\pm} +2\pi p_Y^{\pm} =0, \quad {\dot p}_Y^{\pm} +{\textstyle
{\frac {1}{2\pi}}} \partial_1 \partial_1 Y^{\pm} =0, \nonumber\\[2pt]
&& \partial_{-} \vf^{\pm} =0, \quad  \partial_{-} \c^{\pm} =0. 
\eea
Recalling that we can always stick with the NS periodicity conditions
thanks to spectral flow, we finally write down the general solution 
in the right--moving sector 
\bea\label{cs1}
&&X^{\pm} (\tau, \s)=x^{\pm}+\tau p^{\pm} +{\textstyle{\frac i2}} 
\sum_{n\ne 0} {\textstyle{\frac 1n}} a_n^{\pm} e^{-in(\tau -\s)}
+{\textstyle{\frac i2}} \sum_{n\ne 0} {\textstyle{\frac 1n}} 
b_n^{\pm} e^{-in (\tau +\s)},\nonumber\\[2pt]
&&Y^{\pm} (\tau, \s)=y^{\pm}+\tau p_y^{\pm} +{\textstyle{\frac i2}} 
\sum_{n\ne 0} {\textstyle{\frac 1n}} {\tilde a}_n^{\pm} e^{-in(\tau -\s)}
+{\textstyle{\frac i2}} \sum_{n\ne 0} {\textstyle{\frac 1n}} 
{\tilde b}_n^{\pm} e^{-in (\tau +\s)},\nonumber\\[2pt]
&&\vf^{\pm}={\textstyle{\frac {1}{2\sqrt \pi}}} \sum_{r\in Z+1/2} 
f_r^{\pm} e^{-ir(\tau +\s)}, \quad
\c^{\pm}={\textstyle{\frac {1}{2\sqrt \pi}}} \sum_{r\in Z+1/2} 
g_r^{\pm} e^{-ir(\tau +\s)}. 
\eea
The index ``${+}$'' is not to be confused with the hermitian conjugation 
which in the following  we denote by ``${*}$''. 

The reality condition for the fields together with the 
brackets~(\ref{fermbrac}) and (\ref{br1}) specify the hermiticity properties
and the Poisson brackets for the Fourier modes in the usual way, 
\bea\label{modes}
&& \{ a_n^{-},a_m^{+} \}=in \delta_{n+m,0}, \quad
\{ a_n^{+},a_m^{-} \}=in \delta_{n+m,0}, \quad 
{(a_n^{\pm})}^{*}=a_{-n}^{\pm}, \nonumber\\[2pt]
&& \{ x^{-},p^{+} \}=-{\textstyle{\frac 12}}, \quad \{ x^{+},p^{-} \}=
-{\textstyle{\frac 12}}, \nonumber\\[2pt] 
&& \{ f_r^{+},f_q^{-} \}=-i\delta_{r+q,0}, 
\quad {(f_r^{\pm})}^{*}=f_{-r}^{\pm}.
\eea
The modes $b_n^{\pm}, {\tilde a}_n^{\pm}, {\tilde b}_n^{\pm}, g_r^{\pm}, 
y^{\pm}, p_y^{\pm}$ satisfy precisely the same relations. 

It is noteworthy that the $N{=}2$ superconformal currents given in 
Eq. (\ref{n2}) above are not the maximal closed set one can 
realize on the matter fields. As was pointed out by Siegel~\cite{ws} and
later on by Berkovits and Vafa~\cite{berkvafa}, two more bosonic currents 
(of spin 1 before twisting) and two more fermionic ones (of spin 3/2), 
\be
\epsilon^{ab} \p_a \p_b =0, \quad \epsilon^{ab} \bp_a \bp_b =0,
\quad  \epsilon^{ab} (p_{\bz a} +{\textstyle{\frac {1}{2\pi}}} \partial_1 z_a)
\p_b =0, \quad \epsilon^{ab} (p_{z a} +{\textstyle{\frac {1}{2\pi}}} 
\partial_1 {\bz}_a) {\bp}_b =0,
\ee 
or, in real notation,
\bea\label{car2}
 && \tilde {G_1}=(p^{-} +{\textstyle{\frac {1}{2\pi}}} \partial_1 X^{-}) 
\c^{+}-(p^{+} +{\textstyle{\frac {1}{2\pi}}} \partial_1 X^{+}) \c^{-}-
(p_Y^{-} -{\textstyle{\frac {1}{2\pi}}} 
\partial_1 Y^{-})\vf^{+}+\nonumber\\[2pt]
&& \quad \quad (p_Y^{+} -{\textstyle{\frac {1}{2\pi}}} \partial_1 Y^{+})
\vf^{-}\nonumber\\[2pt]
&& \tilde {H_1}=-(p^{-} +{\textstyle{\frac {1}{2\pi}}} \partial_1 X^{-}) 
\vf^{+}+(p^{+} +{\textstyle{\frac {1}{2\pi}}} \partial_1 X^{+}) \vf^{-}-
(p_Y^{-} -{\textstyle{\frac {1}{2\pi}}} 
\partial_1 Y^{-})\c^{+}+\nonumber\\[2pt]
&& \quad \quad (p_Y^{+} -{\textstyle{\frac {1}{2\pi}}} \partial_1 Y^{+})\c^{-},
\nonumber\\[2pt]
&& \tilde {J_1} =i\c^{-}\vf^{+}-i\c^{+}\vf^{-}=0, \quad     
\tilde {J_2} =i\vf^{+}\vf^{-}-i\c^{+}\c^{-}=0,
\eea
extend the algebra to a ``small'' $N{=}4$ superconformal 
algebra. We remark that the triplet of spin--one currents ($\tilde J,
{\tilde J}_1, {\tilde J}_2$) forms an $su(1,1)$ subalgebra. Furthermore, 
adding the new currents enlarges the symmetry of the formalism.
Besides the $U(1,1)$ group which comes with the $N{=}2$ string currents
and acts as a trivial automorphism on the $N{=}4$ superconformal algebra
just leaving each current invariant, there appears an extra external
automorphism group $U(1,1)$ which maps $\tilde G$, $\tilde H$, $\tilde G_1$,
$\tilde H_1$ generators one into another. Before twisting, the symmetry of the
topological extension is $U(1,1)\times U(1,1)$ which includes the full
Lorentz group $SO(2,2)$. After a topological twist~\footnote{
It is worth recalling that a topological twist by $J$
implemented in Ref.~\cite{berkvafa} does not treat all the currents on 
equal footing. The twisting chooses a $U(1)$ subgroup of the $SU(1,1)$, thus
breaking the full Lorentz group $SO(2,2)$ to $U(1,1)$.}, 
this $N{=}4$ extension 
turns out to be equivalent to the $N{=}2$ formulation, as has been 
demonstrated by the computation of scattering amplitudes~\cite{berkvafa}.
It should be stressed that, being quantum mechanically equivalent to the 
$N{=}2$ string after the twisting, the $N{=}4$ topological formalism offers 
a larger freedom in formulating a heterotic theory. Since prior to the twisting
there are two global $U(1,1)$ groups available, it seems reasonable 
to preserve {\it at least one} when adding
a left--moving sector. In the next section we shall propose the 
left--movers and establish explicit space--time supersymmetry. 
Interestingly enough, we shall find that exactly this type of scenario takes
place and the left movers are compatible with {\it only one} of the $U(1,1)$
groups intrinsic to the $N{=}4$ topological formalism and they explicitly
violate the other. To put it in other words, one could proceed with
the ordinary $N{=}2$ string taken to describe
right movers. These are invariant under $U(1,1)$. 
Then one could add manifestly sypersymmetric left movers
which prove to be invariant under another $U(1,1)$. 
What happens is that each of the $U(1,1)$ groups leaves only one 
chiral half invariant at a time and not the full formalism. A remarkable 
fact, however, is that when acting on the $N{=}2$ superconformal currents, 
the left $U(1,1)$ automatically generates the ``small'' $N{=}4$ superconformal 
algebra. In order to keep a single $U(1,1)$ for the whole string one 
is forced to turn to the topological description by Berkovits and Vafa,
the global symmetry group of the full formalism being the $U(1,1)$
coming with the left movers. We thus conclude that, 
together with the Hamiltonian~(\ref{ham}), the currents~(\ref{car1}) and
(\ref{car2}) completely specify the right--moving sector. 

\vspace{0.6cm}

\noindent
{\bf 3. The left--moving sector}\\[-4pt]

\noindent
Since the string coordinates are automatically decomposed into right and
left modes, they are already present in the left sector.
In the spirit of the GS string, we add space--time spinors which
are world--sheet scalars.
In spite of the heterotic construction adopted in
this paper, we still choose to keep a balance between left-- and 
right--moving degrees of freedom and introduce two canonical pairs,
\be
\{ \t^{(+)}(\s),p_{(+)} (\s') \}=\delta(\s-\s'), \quad
\{ \t^{(-)}(\s),p_{(-)} (\s') \}=\delta(\s-\s'),
\ee 
where $(\t^{(+)},\t^{(-)})$ are real and $(p_{\t (+)},p_{\t (-)})$
are imaginary. The indices $(\pm)$ signify weights equal to $\pm{1\over2}$
with respect to the $SO(1,1)$ subgroup of the full $U(1,1)$ group 
and are viewed as would-be spinor indices.
In this space one immediately observes the intriguing possibility to have 
a $c{=}0$ conformal field theory, matching perfectly the one 
for the right--movers. 

The dynamics of the new fields gets fixed by adding two new terms 
to the Hamiltonian~(\ref{ham}) to arrive at
\bea\label{ham1}
&&H=\int^{2\pi}_{0}\!\! \{ -4\pi p^{+} p^{-}  -4\pi p_Y^{+} p_Y^{-}
-{\textstyle{\frac {1}{\pi}}} \partial_1 X^{+} \partial_1 X^{-}
-{\textstyle{\frac {1}{\pi}}} \partial_1 Y^{+} \partial_1 Y^{-}+
i\vf^{+} \partial_1 \vf^{-}+
\nonumber\\[2pt]
&& \quad \quad  i\vf^{-} \partial_1 \vf^{+}
+i\c^{+} \partial_1 \c^{-} +i\c^{-} \partial_1 \c^{+}
-p_{(+)} \partial_1  \t^{(+)} -p_{(-)} \partial_1 \t^{(-)} \}.
\eea
This yields the free equations of motion 
\be
\partial_{+} \t^{(\pm)} =0, \quad
\partial_{+} p_{\t (\pm)} =0.
\ee
Since the GS-type scalars are single--valued on the world--sheet, 
they must be taken to be periodic functions on the cylinder.

The set of currents we postulate in the sector is
\bea\label{leftcur}
&&T=-2(p^{-} -{\textstyle{\frac {1}{2\pi}}} \partial_1 X^{-})
(p^{+} -{\textstyle{\frac {1}{2\pi}}} \partial_1 X^{+})-2(p_Y^{-} +
{\textstyle{\frac {1}{2\pi}}} \partial_1 Y^{-})
(p_Y^{+} +{\textstyle{\frac {1}{2\pi}}} \partial_1 Y^{+})-\nonumber\\[2pt]
&&\quad \quad {\textstyle{\frac {1}{\pi}}} p_{(+)} \partial_1  \t^{(+)} 
-{\textstyle{\frac {1}{\pi}}} p_{(-)} \partial_1  \t^{(-)}=0, 
\nonumber\\[2pt]
&& G=(p^{-} -{\textstyle{\frac {1}{2\pi}}} \partial_1 X^{-}) p_{(-)} =0,
\quad H=(p_Y^{-} +{\textstyle{\frac {1}{2\pi}}} \partial_1 Y^{-}) p_{(-)} =0,
\nonumber\\[2pt]
&& J =p_{(+)} p_{(-)} =0,     
\eea
satisfying
\be
\partial_{+} T=0, \quad \partial_{+} G=0, \quad \partial_{+} H =0, \quad
\partial_{+} J=0.
\ee
It is relevant to note that the constraints proposed look very similar to
those studied in~\cite{pope1}. They differ, however,  
in the global symmetry structure. The currents studied in~\cite{pope1}
support the full Lorentz group $SO(2,2)$, while in the present formalism
we are forced to stick with a $U(1,1)$ subgroup.

The only non-vanishing brackets in the corresponding algebra are
\bea\label{algebra1}
&&\{ T(\s),T(\s') \}=-{\textstyle{\frac {2}{\pi}}} T(\s) 
\partial_1 \delta(\s-\s') -{\textstyle{\frac {1}{\pi}}} \partial_1 T(\s) 
\delta(\s-\s'),\nonumber\\[2pt]
&&\{ G(\s),T(\s') \}=-{\textstyle{\frac {2}{\pi}}} G(\s) 
\partial_1 \delta(\s-\s') -{\textstyle{\frac {1}{\pi}}} \partial_1 G(\s) 
\delta(\s-\s'),\nonumber\\[2pt]
&&\{ H(\s),T(\s') \}=-{\textstyle{\frac {2}{\pi}}} H(\s) 
\partial_1 \delta(\s-\s') -{\textstyle{\frac {1}{\pi}}} \partial_1 H(\s) 
\delta(\s-\s'),\nonumber\\[2pt]
&&\{ J(\s),T(\s') \}=-{\textstyle{\frac {2}{\pi}}} J(\s) 
\partial_1 \delta(\s-\s') -{\textstyle{\frac {1}{\pi}}} \partial_1 J(\s) 
\delta(\s-\s'),
\eea
implying that all the currents carry conformal spin 2. Two comments are 
in order. Firstly, the set constructed is invariant
under $U(1,1)$ (see Sect.~4 for the explicit realization). Secondly, 
making use of arguments like those exploited in~\cite{pope1} 
one can show that the set above is {\it functionally dependent}
which brings serious problems when one tries to apply BRST techniques.
This may be seen by evaluating the functional determinant of the matrix 
of first derivatives of the constraints with respect to all the variables 
in the problem: its rank equals two. Unfortunately, choosing
an irreducible subset would break the $U(1,1)$ symmetry. Interestingly, this 
seems to be analogous to the situation in the conventional $d{=}10$ 
Green--Schwarz superstring.\footnote{
The implication of computing conventional constraints in
terms of dual fields for the type-II Green-Schwarz theory has been
considered in \cite{b3}.
For a detailed discussion about the covariant quantization 
and kappa symmetry, see e.g. \cite{b1}.}

Let us turn to the issue of global supersymmetry. Since, by the
very construction,  the subspace spanned by $(X^{+},\t^{(+)})$ is nothing 
other than a $(1,0)$--superspace (see Ref.~\cite{dg} for our definition), 
it is straightforward to realize the following {\it on--shell} global
supersymmetry, 
\bea
&&\delta \t^{(+)} =\e^{(+)}, \quad \delta X^{+} = i\t^{(+)} \e^{(+)},
\nonumber\\[2pt]
&&\delta p_{(+)} =-2i(p^{-} -{\textstyle{\frac {1}{2\pi}}} 
\partial_1 X^{-}) \e^{(+)}, \quad \delta p^{+} =
-{\textstyle{\frac {i}{2\pi}}} \partial_1 \t^{(+)} \e^{(+)}.
\eea
As usual, the corresponding current is defined via the Poisson bracket,
\be
\delta A =\{ A, \int^{2\pi}_{0}\!\! d \s' q_{(+)} (\s') \} \e^{(+)}, 
\ee
which yields
\bea\label{supercurrent}
&& q_{(+)} = p_{(+)} -2i \t^{(+)} (p^{-} -{\textstyle{\frac {1}{2\pi}}} 
\partial_1 X^{-}), \quad
\{ q_{(+)} (\s), q_{(+)} (\s') \} = 2i P^{-} \delta(\s-\s'),
\eea
with $P^{-} \equiv -2 (p^{-} -{\textstyle{\frac {1}{2\pi}}} 
\partial_1 X^{-})$ being the generator of translations in the 
$X^{+}$--direction.

Due to the equation
\be
\partial_{+} q_{(+)}=0,
\ee
the integral
\be
Q_{(+)} = \int^{2\pi}_{0}\!\! d \s q_{(+)} (\s)
\ee
gives a conserved charge. The supersymmetry 
transformations defined above leave the constraint surface 
(strongly) invariant, as they should.

Finally, we give the classical solutions in the sector (recall that 
$p_{\t (+)}$ and $p_{\t (-)}$ are imaginary):
\bea\label{fermleft}
&&\t^{(\pm)}={\textstyle{\frac {1}{\sqrt {2\pi}}}} \sum_{n\in Z} 
d_n^{(\pm)} e^{-in(\tau -\s)}, \quad
p_{(\pm)}={\textstyle{\frac {i}{\sqrt {2\pi}}}} \sum_{n\in Z} 
\ell_{n (\pm)} e^{-in(\tau -\s)}, \nonumber\\[2pt]
&& \{ d_n^{(\pm)},\ell_{m (\pm)} \}=-i\delta_{n+m,0}, \quad 
{(d_n^{(\pm)})}^{*}=d_{-n}^{(\pm)}, \quad  
{(\ell_{n (\pm)})}^{*}=\ell_{-n (\pm)}.
\eea
To conclude the section, it is worth noting that the structure of the 
left--moving ``spinors'' resembles what one usually expects of a ghost system. 
Each pair $(\t^{(+)},p_{(+)})$, $(\t^{(-)},p_{(-)})$ makes a contribution 
of $-2$ to the conformal anomaly, and altogether they cancel $c{=}4$ 
coming from the  $X^{\pm},Y^{\pm}$ matter system. Also one suspects them to
bring about a degeneracy of the vacuum. We shall turn to this 
issue later in Sect.~6, where we shall make explicit use of this degeneracy in 
establishing global supersymmetry of the string spectrum.

\vspace{0.6cm}

\noindent
{\bf 4. Global symmetry structure}\\[-4pt]

\noindent
Having formulated both left and right--movers, we are now in a position to 
discuss the global symmetry structure of the full theory. For 
the right--movers it is trivial to transform the common $U(1,1)$ 
transformations to the real notation. We gather them in the following table.

\vskip 0.2cm
\begin{center}
\begin{tabular}{|l|c|c|c|c|c|c|c|c|c| }
\hline \vphantom{$\displaystyle\int$}
& $X^{+}$ &  $X^{-}$ & $Y^{+}$ & $Y^{-}$ & $\vf^{+}$ & $\vf^{-}$ & 
$\c^{+}$ & $\c^{-}$  \\
\hline \vphantom{$\displaystyle\int$}
$\delta_\a$ & $\a X^{+}$ &  $-\a X^{-}$ & $\a Y^{+}$ & $-\a Y^{-}$ & 
$\a \vf^{+}$ & $-\a \vf^{-}$ & $\a \c^{+}$ & $-\a \c^{-}$\\
\hline \vphantom{$\displaystyle\int$}
$\delta_\b$ & $\b Y^{+}$ &  $\b Y^{-}$ & $-\b X^{+}$ & $-\b X^{-}$ & 
$\b \c^{+}$ & $\b \c^{-}$ & $-\b \vf^{+}$ & $-\b \vf^{-}$\\
\hline \vphantom{$\displaystyle\int$}
$\delta_{\g^{++}}$ & $\g^{++} Y^{-}$ & 0 & $-\g^{++} X^{-}$ & 0 & 
$\g^{++} \c^{-}$ & 0 & $-\g^{++} \vf^{-}$ & 0 \\
\hline \vphantom{$\displaystyle\int$}
$\delta_{\g^{--}}$ & 0 &  $\g^{--} Y^{+}$ & 0 & $-\g^{--} X^{+}$ & 0 & 
$\g^{--} \c^{+}$ 
& 0 & $-\g^{--} \vf^{+}$\\
\hline
\end{tabular}
\end{center}
\begin{center}
Table 1. Global U(1,1) transformations acting in the right sector.
\end{center}

\vskip 0.2cm

These leave invariant the equations of motion and map the
$N{=}2$ superconformal currents into each other. The corresponding conserved
charges, obviously, form a $u(1,1)$ algebra. Being symmetries of the right 
sector, the transformations, however, do not hold in the left sector as 
they do not leave the currents~(\ref{leftcur}) invariant. However, in the 
left sector one finds an independent $u(1,1)$ realized in the following way.
\vskip 0.2cm
\begin{center}
\begin{tabular}{|l|c|c|c|c|c|c|c|c|c| }
\hline \vphantom{$\displaystyle\int$}
& $X^{+}$ &  $X^{-}$ & $Y^{+}$ & $Y^{-}$ & $\t^{(+)}$ & $p_{(+)}$ & 
$\t^{(-)}$ & $p_{(-)}$  \\
\hline \vphantom{$\displaystyle\int$}
$\delta_\o$ & $\o X^{+}$ &  $-\o X^{-}$ & 0 & 0  & 
$ {\textstyle{\frac 12}} \o \t^{(+)}$ & $ -{\textstyle{\frac 12}} \o 
p_{(+)}$ &  $ -{\textstyle{\frac 12}} \o \t^{(-)}$ &
$ {\textstyle{\frac 12}} \o p_{(-)}$\\    
\hline \vphantom{$\displaystyle\int$}
$\delta_\z$ & 0 & 0 & $\z Y^{+}$ &  $-\z Y^{-}$ & 
$ {\textstyle{\frac 12}} \z \t^{(+)}$ & $ -{\textstyle{\frac 12}} \z 
p_{(+)}$ &  $ -{\textstyle{\frac 12}} \z \t^{(-)}$ &
$ {\textstyle{\frac 12}} \z p_{(-)}$\\    
\hline \vphantom{$\displaystyle\int$}
$\delta_\l$ & 0 & $\l Y^{-}$ & $-\l X^{+}$ & 0 & 0 & 0 & 0 & 0 \\
\hline \vphantom{$\displaystyle\int$}
$\delta_\m$ & $\m Y^{+}$ & 0 & 0 & $-\m X^{-}$ & 0 & 0 & 0 & 0 \\
\hline
\end{tabular}
\end{center}
\begin{center}
Table 2. Global U(1,1) transformations acting in the left sector.
\end{center}

\vskip 0.2cm
We indeed verify that the associated generators, 
\bea
A_1 = -2p^{-} X^{+} +2p^{+} X^{-} +{\textstyle{\frac 12}} p_{(+)} \t^{(+)}
-{\textstyle{\frac 12}} p_{(-)} \t^{(-)},\nonumber\\[2pt]
A_2 = 2p_Y^{-} Y^{+} -2p_Y^{+} Y^{-} +{\textstyle{\frac 12}} 
p_{(+)} \t^{(+)}-{\textstyle{\frac 12}} p_{(-)} \t^{(-)},\nonumber\\[2pt]
A_3 =-2p^{+} Y^{-} -2p_Y^{-} X^{+}, \quad 
A_4 =-2p^{-} Y^{+} -2p_Y^{+} X^{-},
\eea
lead to the conserved charges
\bea\label{concharge}
&& L_1 =-i \int d \s (A_1+A_2), \quad L_2=-i \int d \s (A_1 -A_2), 
\nonumber\\[2pt] 
&& L_3 =-i \int d \s (A_3 +A_4), \quad L_4 =-i \int d \s (A_3 -A_4),
\eea
which obey a $u(1,1)$ algebra. Like the transformations discussed above,
these symmetries do not hold for the other sector. The crucial observation,
however, is that extending the charge $L_1$ by two new contributions (which 
do not spoil the algebra!),
\be
L_1 \rightarrow {L'}_1=-i \int d \s (A_1+A_2 +2i\vf^{-}\vf^{+} + 
2i\c^{-}\c^{+}),
\ee
one arrives at the transformations which map the currents $\tilde G, \tilde H$ 
of the right sector onto $\tilde {G_1}, \tilde {H_1}$  and vice versa.
Then the closure of the algebra automatically produces the remaining
currents $\tilde {J_1}, \tilde {J_2}$ of the $N{=}4$ topological formalism.

Thus, we see that by adopting the $N{=}4$ topological description for the 
right--movers, one can have a unique $U(1,1)$ global group acting on 
the right-- and left--movers simultaneously while keeping space--time 
supersymmetry in the left sector. It should be stressed that, although
there are two $U(1,1)$ groups intrinsic to the $N{=}4$ formalism, only
one of them is compatible with the left--moving sector, while the other
is explicitly violated by the latter. The global symmetry group of the 
whole string is thus a single $U(1,1)$. 

In the previous section, the global supersymmetry transformation was 
specified. Adjoining it to the $U(1,1)$ transformations, the larger algebra
closes upon adding two new transformations with the fermionic parameters
$\r^{(+)}$ and $\k^{(+)}$,
\bea
&&\d_{\r} Y^{+} = i\t^{(+)} \r^{(+)}, \quad 
\d_{\r} p_Y^{+} ={\textstyle{\frac {i}{2\pi}}} \partial_1 
\t^{(+)} \r^{(+)}, \quad \d_{\r} p_{(+)} =2i(p_Y^{-} +
{\textstyle{\frac {1}{2\pi}}} \partial_1 Y^{-}) \r^{(+)}, \nonumber\\[2pt] 
&& \d_{\r} \t^{(+)}=0; \quad \quad \d_{\k} \t^{(+)}=\k^{(+)}.
\eea
It has to be mentioned that these leave invariant the currents in the 
right sector. The corresponding conserved charges are given by
\be
S_{(+)} = 2i \int d \s (p_Y^{-} +{\textstyle{\frac {1}{2\pi}}} \partial_1 
Y^{-}) \t^{(+)}, \quad {\tilde S}_{(+)} = \int d \s p_{(+)}. 
\ee
Taking into account the additional charges associated with the translation 
invariance $\d X^\pm =a^\pm$ and $\d Y^\pm =b^\pm$,
\bea
B^{\pm} = -2 \int d \s (p^{\pm} -{\textstyle{\frac {1}{2\pi}}} 
\partial_1 X^{\pm}),\quad B_Y^{\pm} = 2\int d \s (p_Y^{\pm} +
{\textstyle{\frac {1}{2\pi}}} \partial_1 Y^{\pm}),
\eea
one can finally write down a complete superalgebra. Most compactly, this 
is represented by the following table (equal--time brackets involving 
$B^{\pm}$, $B_Y^{\pm}$ vanish and are omitted here):
\vskip 0.2cm

\begin{center}
\begin{tabular}{|l|c|c|c|c|c|c|c|c|c|c|c|c| }
\hline \vphantom{$\displaystyle\int$}
\{, \}& Q & S & $\tilde S$ & $L_1$ & $L_2$ & $L_3$ & 
$L_4$ \\
\hline \vphantom{$\displaystyle\int$}
Q & $2iB^{-}$ &  $iB_Y^{-}$ & $iB^{-}$ & $i\tilde S$ & 
$i(Q-\tilde S)$ & $-iS$ & $-iS$ \\
\hline \vphantom{$\displaystyle\int$}
S & $iB_Y^{-}$ & 0 &  $iB_Y^{-}$ & 0 & $-iS$ & $i(Q-\tilde S)$ & 
$ -i(Q-\tilde S)$ \\ 
\hline \vphantom{$\displaystyle\int$} 
$\tilde S$ & $iB^{-}$ & $iB_Y^{-}$ & 0 & $i\tilde S$ & 0 & 0 & 
0 \\
\hline \vphantom{$\displaystyle\int$} 
$L_1$ & $-i\tilde S$ & 0 & $-i\tilde S$ & 0 & 0 & 0 & 
0 \\
\hline \vphantom{$\displaystyle\int$} 
$L_2$ & $-i(Q-\tilde S)$ & $iS$ & 0 & 0 & 0 & $2iL_4$ & 
$2iL_3$ \\
\hline \vphantom{$\displaystyle\int$} 
$L_3$ & $iS$ & $-i(Q-\tilde S)$ & 0 & 0 & $-2iL_4$ & 0 
& $2iL_2$ \\
\hline \vphantom{$\displaystyle\int$} 
$L_4$ & $iS$ & $i(Q-\tilde S)$ & 0 & 0 & $-2iL_3$ & 
$-2iL_2$ & 0 \\
\hline  \vphantom{$\displaystyle\int$} 
$B^{+}$ & 0 & 0 & 0 & $-iB^{+}$ & $-iB^{+}$ & 
$-iB_Y^{+}$ & $iB_Y^{+}$ \\
\hline \vphantom{$\displaystyle\int$} 
$B^{-}$ & 0 & 0 & 0 & $iB^{-}$ & $iB^{-}$ & 
$-iB_Y^{-}$ & $-iB_Y^{-}$ \\
\hline \vphantom{$\displaystyle\int$} 
$B_Y^{+}$ & 0 & 0 & 0 & $-iB_Y^{+}$ & $iB_Y^{+}$ & 
$iB^{+}$ & $iB^{+}$ \\
\hline \vphantom{$\displaystyle\int$} 
$B_Y^{-}$ & 0 & 0 & 0 & $iB_Y^{-}$ & $-iB_Y^{-}$ & 
$iB^{-}$ & $-iB^{-}$ \\
\hline  
\end{tabular}
\end{center}
\begin{center}
Table 3. The supersymmetry algebra of the full model.
\end{center}

\vskip 0.2cm

It is straightforward to verify that the Jacobi identities hold for this
superalgebra. Note further that the bracket of the $S_{(+)}$ charge with 
itself is zero. The same is true for ${\tilde S}_{(+)}$. 
Because the generators are
composed of ``spinors'', at the quantum level they will be represented by 
nilpotent operators, failing the usual unitarity argument to show that
these symmetries are trivial. In Sect.~6 we will see that  
$S_{(+)},{\tilde S}_{(+)}$ indeed act on a quantum space, although not 
playing a significant role. 
 
Before closing this section, it seems instructive to inspect the above
symmetry transformations in complex notation. Putting $z^a$ and 
${\bar z}^a$ in a single row  $Z^A =(z^a,{\bar z}^a)$ with $A{=}1,2,3,4$,
one can conveniently rewrite the (infinitesimal) transformations gathered 
in Table~1 as follows,
\be
\delta Z^A=i\a^i {L_i^A}_B Z^B, \qquad {\bar L_i}^{T} \eta =\eta L_i,
\qquad \eta_{AB}={\textstyle{diag}}(-,+,-,+),
\ee 
where the matrices $L_i$ form a basis for the $u(1,1)$ algebra, 
$$
\begin{array}{lll}
&& L_1=\left(\begin{array}{cccc} 
0 & -i & 0 & 0\\
-i & 0 & 0 & 0\\
0 & 0 & 0 & -i\\
0 & 0 & -i & 0\
\end{array}\right), \qquad \qquad 

L_2=\left(\begin{array}{cccc} 
-1 & 0 & 0 & 0\\
0 & -1 & 0 & 0\\
0 & 0 & +1 & 0\\
0 & 0 & 0 & +1\
\end{array}\right), \nonumber\\[2pt]

\vspace{0.5cm} 

&& L_3=
\left(\begin{array}{cccc} 
-1 & +1 & 0 & 0\\
-1 & +1 & 0 & 0\\
0 & 0 & +1 & -1\\
0 & 0 & +1 & -1\
\end{array}\right), \qquad \qquad

L_4=
\left(\begin{array}{cccc} 
-1 & -1 & 0 & 0\\
+1 & +1 & 0 & 0\\
0 & 0 & +1 & +1\\
0 & 0 & -1 & -1\
\end{array}\right). 
\end{array}
\eqno{(43)}
$$
\addtocounter{equation}{1}
Analogously, for the transformations from Table 2 one finds another set 
of basis elements,
$$
\begin{array}{lll}
&& L_1=\left(\begin{array}{cccc} 
0 & -i & 0 & 0\\
-i & 0 & 0 & 0\\
0 & 0 & 0 & -i\\
0 & 0 & -i & 0\
\end{array}\right), \qquad \qquad 

L_2=\left(\begin{array}{cccc} 
0 & 0 & 0 & -i\\
0 & 0 & -i & 0\\
0 & -i & 0 & 0\\
-i & 0 & 0 & 0\
\end{array}\right), \nonumber\\[2pt]

\vspace{0.5cm} 

&& L_3=\left(\begin{array}{cccc} 
-1 & 0 & 0 & 0\\
0  & -1 & 0 & 0\\
0 & 0 & +1 & 0\\
0 & 0 & 0 & +1\
\end{array}\right), \qquad \qquad

L_4=\left(\begin{array}{cccc} 
0 & 0 & 0 & -1\\
0 & 0 & -1 & 0\\
0 & +1 & 0 & 0\\
+1 & 0 & 0 & 0\
\end{array}\right). 
\end{array}
\eqno{(44)}
$$
\addtocounter{equation}{1}
We see from here that in the former case the matrices are block--diagonal.
In other words, the fields $z^a$ and ${\bar z}^a$ do not mix 
under these transformations. In the latter case, some of the 
generators are off--diagonal, mixing $z^a$ with ${\bar z}^a$. This 
means, in particular, that a combination of the form ${\bar z}^a \eta_{ab} 
y^b$, which is trivially invariant under the action of the right 
$U(1,1)$, does not hold invariant under the action of the left $U(1,1)$, 
and a more general object like ${\bar Z}^A \eta_{AB} Y^B$ is to be handled 
with. As we shall discuss in Sect.~7, this causes certain problems in
constructing scattering amplitudes because it prevents one from a naive use
of the vertex operator known for the conventional $N{=}2$ string.

\vspace{0.6cm}

\noindent
{\bf 5. Quantized right--movers}\\[-4pt]

\noindent
As has been discussed in Sect.~2, the right--moving sector of our model
relies on the $N{=}4$ topological extension of (a chiral half of) the
$N{=}2$ string. At the level of string amplitudes a detailed proof 
of equivalence of the two formalisms was given in~\cite{berkvafa,bvw}. 
Alternatively, one could proceed directly from 
the small $N{=}4$ superconformal algebra to verify that the positive 
(half--integer) modes of ${\tilde G}_1$, ${\tilde H}_1$ 
kill all physical states, provided so do the zero modes 
of ${\tilde J}_1$ and ${\tilde J}_2$ (see also the discussion in 
Ref.~\cite{ws}). Since even for the smaller $N{=}2$
superconformal algebra a proper analysis shows the absence of excited 
states and because the zero modes of ${\tilde J}_1$ and ${\tilde J}_2$
annihilate the ground state, one arrives at the same spectrum 
as for the ordinary $N{=}2$ string. Below we briefly sketch the main 
points (conformed to our notation).  A more detailed exposition can be 
found in~\cite{ov,berkvafa}.

Given the matter system $X^{\pm},Y^{\pm},\vf^{\pm},\c^{\pm}$ one finds
$c{=}6$ for the conformal anomaly. Introducing the superconformal 
ghosts $(c,b),(\g,\b),(\g_1,\b_1),(c_1, b_1)$ associated with the $N{=}2$ 
superconformal currents $\tilde T,\tilde G,\tilde H,\tilde J$, 
the ghost contribution to the anomaly adds up to $-26+11+11-2=-6$. Thus,  
this sector is critical in $2{+}2$ dimensions. Evaluating the normal-ordering
constants one finds a vanishing critical intercept so that the ground state
is a massless scalar. The BRST analysis shows that the latter is in fact
the only physical state in the spectrum~\cite{bienkowska,ov}.
It is created from the (zero-momentum) NS vacuum by application of a
vertex operator whose form depends on the superconformal ghost picture
\cite{fms} chosen. Abbreviating $K^{\pm} =k^{\pm} +i\kappa^{\pm}$
and suppressing the ghost structure, one has in the $(-1)$ picture
\bea\label{asympt}
&& V_{-1}(K,z) =\ :e^{i(K\bar z + \bar K z)}: = 
:e^{-2i(k^{+} X^{-}+k^{-} X^{+} +\kappa^{+} Y^{-} + 
\kappa^{-} Y^{+})}:, 
\eea
while the $0$-picture vertex operator reads
(conformed to our notation)
\bea\label{vertex}
V_0(K,z) =\ 
:2(iK \partial \bar z -i \bar K \partial z -2\pi \bar K \psi K \bar \psi)\,
e^{i(K\bar z + \bar K z)}:.
\eea

Further analysis shows that only the three--point function does not vanish 
at tree level.  
In the zero--instanton sector one finds\footnote{
The three--point functions in the other instanton sectors are proportional
to this one.}
\bea\label{ampl}
A^{tree}_{right}(1,2,3) &=&  
\langle V_{-1}(K_1,\infty) V_0(K_2,1) V_{-1} (K_3,0) \rangle \nonumber\\[2pt]
&=& K_2 {\bar K}_3 - {\bar K}_2 K_3 \nonumber\\[2pt] 
&=& 2i({k_2}^{+} {\kappa_3}^{-} + 
{k_2}^{-} {\kappa_3}^{+})-2i({k_3}^{+} {\kappa_2}^{-} + 
{k_3}^{-} {\kappa_2}^{+}).
\eea
Due to the relation between open- and closed--string amplitudes~\cite{kaw1}, 
one might expect that Eq.~(\ref{ampl})
is the only contribution to the full S-matrix coming from the right 
sector. However, the story is not so simple because, upon a more careful
inspection, one observes that Eq.~(\ref{vertex}) is not 
invariant under the global $U(1,1)$ group installed by the 
left--movers. In other words, disregarding the full set of $N{=}4$ 
superconformal currents and constructing the interactions by taking into 
account only the $N{=}2$ subset breaks the manifest $U(1,1)$ covariance of 
the full formalism. We will return to this issue in Sect.~7.

\vspace{0.6cm}

\noindent
{\bf 6. Quantized left--movers}\\[-4pt]

\noindent
As we have seen in Sect.~3, the
constraints intrinsic to the left--movers are (infinitely) reducible. 
This entails a serious complication for the BRST 
procedure (for a similar point see e.g.~\cite{pope1}).
For this reason we employ the conventional operator method
for the covariant quantization.

Because bosonic fields are common for both right and left sectors,
it suffices to discuss the ``spinors''. 
The zero-mode algebra\footnote{
{}From now on, $[,]$ and $\{,\}$ denote commutators and anticommutators,
respectively.}
\be
\{d_0,\ell_0\}=1 ,\quad {d_0}^2=0={\ell_0}^2
\ee
enforces a two--fold degeneracy of the vacuum.
The fields $\t^{(-)}$ and $p_{(-)}$ are handled in precisely the same manner,
so the full vacuum is four--fold degenerate.
In what follows, we discuss in detail a representation for $\t^{(+)}, p_{(+)}$ 
only and for brevity omit the ${\textstyle{(+)}}$ index: 
\bea
&& \ell_0  \arrowvert \downarrow  \rangle =0, \quad  
\ell_0 \arrowvert \uparrow  \rangle =  \arrowvert \downarrow  \rangle, 
\quad d_0  \arrowvert \downarrow \rangle =\arrowvert \uparrow  \rangle, 
\quad d_0  \arrowvert \uparrow  \rangle =0, 
\nonumber\\[2pt]
&& \ell_n   \arrowvert \downarrow  \rangle =\ell_n  
\arrowvert \uparrow  \rangle =d_n  \arrowvert \downarrow  \rangle =
d_n  \arrowvert \uparrow  \rangle =0 \qquad \textrm{for} \quad n{\ge} 1.
\eea
Since we are interested in a unitary representation 
and $d_0^\dagger{=}d_0$, $\ell_0^\dagger{=}\ell_0$, the scalar products are
\bea\label{norms}
\langle \downarrow | \downarrow \rangle =0, \qquad  
\langle \uparrow | \uparrow \rangle =0, \qquad
\langle \downarrow | \uparrow \rangle =
\langle \downarrow | \,\theta\, | \downarrow \rangle =1.
\eea
These relations, as well as the value of the conformal spin carried by the 
``spinors'', identify $(\theta,p)$ as a fermionic $c{=}{-}2$ system.
The latter arises as R~symmetry ghosts in the conventional $N{=}2$ 
string~\cite{bischoff} and as auxiliary fermions in the bosonization 
of the superconformal ghosts of the $N{=}1$ NSR string~\cite{fms}.

As for the right--movers, the normal--ordering constants add to zero.
The quantum version of the algebra~(\ref{algebra1}),
\bea
&& [ L_n,L_m ] =(n-m) L_{n+m},\nonumber\\[2pt]
&& [ L_n,G_m ] =(n-m) G_{n+m},\nonumber\\[2pt]
&& [ L_n,H_m ] =(n-m) H_{n+m},\nonumber\\[2pt]
&& [ L_n,J_m ] =(n-m) J_{n+m},
\eea
implies that physical states in the complete Hilbert space are 
defined by 
\be\label{physical}
L_n|\textrm{phys}\rangle =
G_n|\textrm{phys}\rangle =
H_n|\textrm{phys}\rangle =
J_n|\textrm{phys}\rangle = 0
\qquad \textrm{for}\quad n{\ge}0.
\ee

Because, by the very construction,  there 
are no excited physical states in the right sector, it suffices to 
concentrate on the left--sector ground state. As we have seen 
above this is four--fold degenerate,
\be\label{states}
\Phi (k,\kappa)\, | \downarrow \rangle {\times} | \downarrow \rangle, \quad  
\Psi_{(+)} (k,\kappa)\, | \uparrow \rangle {\times} | \downarrow \rangle, \quad 
\Upsilon_{(-)}(k,\kappa)\,|\downarrow\rangle{\times}|\uparrow\rangle,\quad 
\Sigma (k,\kappa)\, | \uparrow \rangle {\times} | \uparrow \rangle.
\ee
Recall that the operators $d_0^{(\pm)}$,
which raise $|\downarrow\rangle\to|\uparrow\rangle$,
carry weights of $\pm{1\over2}$ with respect to the
$SO(1,1)$ subgroup of the full $U(1,1)$ group and, hence, assign those to the 
excited states ${| \uparrow \rangle}^{(\pm)}$. For brevity we omit 
indices carried by the states throughout the text.

On the ground states~(\ref{states}), the conditions~(\ref{physical})
reduce to their zero--mode part, with
\be
L_0 \to -2(p^-p^+ + p_y^-p_y^+), \quad
G_0 \to p^- \ell_{0(-)}, \quad
H_0 \to p_y^- \ell_{0(-)}, \quad
J_0 \to \ell_{0(+)} \ell_{0(-)}.
\ee
As usual, $L_0$ forces the states to lie on the mass shell
\be
k^{+}k^{-} + \kappa^{+} \kappa^{-} =0. 
\ee
The other three operators are nilpotent. 
The $J_0$ condition eliminates~$\Sigma$.
The remaining constraints $G_0$ and $H_0$ imply
\be
k^{-}\Upsilon_{(-)}=\kappa^{-}\Upsilon_{(-)}=0
\ee
which, for generic kinematics (normalizability of states), 
renders $\Upsilon_{(-)}$ unphysical as well.

Hence, we are left with the pair $(\Phi,\Psi_{(+)})$ for the time being.
A relevant point to check is whether these two remaining massless states 
are indeed connected by supersymmetry. 
Taking the Laurent expansion of the supersymmetry current 
$q_{(+)}$~(\ref{supercurrent}) in the form
\be
q_{(+)}(z) ={\textstyle{\frac {1}{\sqrt {2\pi^2}}}} \sum_{n\in Z} 
\frac {q_n }{{\bar z}^{n+1}}
\ee 
and evaluating the operator products with the superconformal 
currents~(\ref{leftcur}), one easily finds the commutation relations 
\bea
&& [L_n, q_m]=-m q_{n+m}, \qquad \{G_n,q_m \}=0, \nonumber\\[2pt]
&& \{H_n, q_m\} =0, \qquad [J_n, q_m]=-{\textstyle{\frac 
{i {\sqrt {2}}}{\pi}}}G_{n+m}. 
\eea
One sees from here that only the zero mode of $q_{(+)}$, namely
\be
q_{0 (+)} =i {\sqrt \pi }\ell_{0 (+)} -{\textstyle{\frac 
{i}{\sqrt \pi}}} \sum_{n\in Z} d_n^{(+)} a_{-n}^{-} ,
\ee
(weakly) commutes with the super Virasoro generators and, hence, maps 
physical states to physical states. For the ground states at hand this yields
\be
\delta_\textrm{susy} \Phi =i{\sqrt \pi}\epsilon^{(+)}\Psi_{(+)}, \qquad
\delta_\textrm{susy} \Psi_{(+)}=-{\textstyle{\frac 
{i}{\sqrt \pi}}} \epsilon^{(+)} p^{-}\Phi,
\ee
and the corresponding fields do fall into a supermultiplet. 

Let us finally discuss the way the $U(1,1)$ symmetry acts on the physical 
states or, equivalently, on the vertex operators. Since the analysis reduces 
to the zero-mode sector, one can readily deduce from~(\ref{concharge})
a quantum representation for the $u(1,1)$ algebra (the ordering of the 
operators involved is fixed as given below)
\bea
&& L_1= 2x^{+} p^{-}  -2x^{-} p^{+} + 2y^{+} p_y^{-}  -2y^{-} p_y^{+} 
+i d_0^{(+)} \ell_{0 (+)} -i d_0^{(-)} \ell_{0 (-)}, \nonumber\\[2pt]
&& L_2= 2x^{+} p^{-}  -2x^{-} p^{+} - 2y^{+} p_y^{-}  +2y^{-} p_y^{+},
\nonumber\\[2pt]
&& L_3= 2y^{-} p^{+}  -2x^{+} p_y^{-} + 2y^{+} p^{-}  -2x^{-} p_y^{+},
\nonumber\\[2pt]  
&& L_4= 2y^{-} p^{+}  -2x^{+} p_y^{-} - 2y^{+} p^{-}  +2x^{-} p_y^{+},
\eea
were $\{ d_0, \ell_0 \}=1$, $[x^{\pm}, p^{\mp}]=-{\textstyle{\frac {i}{2}}}$,
and $[y^{\pm}, p_y^{\mp}]=-{\textstyle{\frac {i}{2}}}$.
Notice that the operators generating the algebra are hermitian.
One sees from the relations above that the generators $L_2$, $L_3$, $L_4$ 
forming the $su(1,1)$ subalgebra (see Table 3 for the commutation relations) 
do not involve any spin part. This means that the physical states are 
singlets with respect to the $SU(1,1)$ subgroup of the full $U(1,1)$. 
The generator $L_1$ of the abelian subalgebra $u(1)$, however, contains
a nontrivial ``spin'' part $d_0^{(+)} \ell_{0 (+)}$ which
measures the helicity of our massless states $\Phi$ and $\Psi_{(+)}$
as $0$ and $1/2$, respectively, and thus distinguishes between them. 
In contrast to the ordinary $N{=}2$ string, where states of different
helicity are physically equivalent due to the action of picture-raising
and spectral flow, $\Phi$ and $\Psi_{(+)}$ represent inequivalent states
because of their left-moving structure.

\vspace{0.6cm}

\noindent
{\bf 7. Discussion}\\[-4pt]

\noindent
To summarize, in the present paper we examined a novel possibility
for including space--time fermions into the $N{=}2$ string. In contrast to
previous approaches we chose to maintain the $U(1,1)$ covariance intrinsic 
to the original $N{=}2$ string and supersymmetrized only one direction 
of the configuration space. Supersymmetry has been realized explicitly
on the left--movers. Interestingly enough, application of the $U(1,1)$ 
transformations to the standard set of $N{=}2$ superconformal currents,
which was our starting point in the right--moving sector, automatically 
generates the $N{=}4$ currents of the topological reformulation by
Berkovits and Vafa. It seems that the space--time supersymmetry distinguishes 
the latter from the former. Quantization has been accomplished, yielding two 
physical states in the spectrum, one boson and one fermion, which form 
a multiplet of the global $(1,0)$ supersymmetry.  

We turn to discuss some points which have not been covered in the paper and
may constitute further developments. Firstly, since from the outset
Lorentz covariance has been sacrificed, the question of spin carried by
the quantum states in the spectrum can be answered only after an explicit 
evaluation of tree--level scattering amplitudes. However, this immediately
reveals a difficulty because the (right--sector) vertex operator~(\ref{vertex})
does not respect the $U(1,1)$ covariance of the full formalism
which is kept by the left--movers (see Sect.~4). 
Secondly, the computation of scattering amplitudes requires a determination
of the $(\theta,p_\theta)$ zero--mode measure in the vertex operator
correlation functions. As is familiar from ghost systems, 
such zero--mode insertions can be subsumed in a modification of the 
scalar product~(\ref{norms}) plus a change of the conjugation rules.
Also, the full vertex operators~$V_0$ are yet to be constructed. 
These questions require a further investigation and we hope to 
report on them elsewhere. Thirdly, throughout the paper we worked in the 
superconformal gauge. A Lagrangian formulation is missing and suggests 
another interesting problem. Yet, from the structure of the constraints
one may already suspect that a Lagrangian formulation is likely
to possess both kappa symmetry (acting on the left--moving fields) as well as 
an $N{=}2$ world--sheet local supersymmetry (acting on the right--movers). 
This could provide an interesting combination of Green--Schwarz and 
Neveu--Schwarz--Ramond formalisms.

\newpage

\noindent
{\bf Acknowledgments}\\[-4pt]

\noindent
The work of two of us (S.B. and A.G.) has been supported by the Iniziativa 
Specifica MI12 of the Commissione IV of INFN and by INTAS grant No
00 OPEN 254.  A.G. thanks the ITP at Hannover
University for the hospitality extended to him.
The third author (O.L.) is partially supported by DFG under grant Le~838/7-1.
He also acknowledges hospitality of the Laboratori Nazionali di Frascati.

\end{document}